\documentclass[preprint]{aastex6}
\usepackage{amsmath, amsthm,amssymb}
\usepackage{multirow}
\usepackage{url}

\usepackage{dcolumn}
\newcolumntype{d}[1]{D{.}{.}{#1}}

\begin{document}

\title{Phoebe: a surface dominated by water}

\author{Wesley C. Fraser \altaffilmark{1} and Michael E. Brown \altaffilmark{2}}
\affil{1 - Queen's University, Belfast\\ Belfast Co. Antrim\\ UK BT7 1NN}
\email{1 - wes.fraser@qub.ac.uk}
\affil{2 - California Institute of Technology}

\date{} % delete this line to display the current date

\slugcomment{Submitted to ApJ}
\received{?}
\revised{?}

\begin{abstract}
The Saturnian irregular satellite, Phoebe, can be broadly described as a water-rich rock. This object, which presumably originated from the same primordial population shared by the dynamically excited Kuiper Belt Objects, has received high resolution spectral imaging during the Cassini flyby. We present a new analysis of the Visual Infrared Mapping Spectrometer observations of Phoebe, which critically, includes a geometry correction routine that enables pixel-by-pixel mapping of visible and infrared spectral cubes directly onto the Phoebe shape model, even when an image exhibits significant trailing errors. The result of our re-analysis is a successful match of 40 images, producing spectral maps covering the majority of Phoebe's surface, roughly a 3rd of which is imaged by high resolution observations ($<22$~km per pixel resolution). There is no spot on Phoebe's surface that is absent of water absorption. The regions richest in water are clearly associated with the Jason and South Pole impact basins.  We find Phoebe exhibits only three spectral types, and a water-ice concentration that correlates with physical depth and visible albedo. The water-rich and water-poor regions exhibit significantly different crater size frequency distributions, and different large crater morphologies. We propose that Phoebe once had a water-poor surface whose water-ice concentration was enhanced by basin forming impacts which exposed richer subsurface layers. Finally, we demonstrate that the range of Phoebe's water-ice absorption spans the same range exhibited by dynamically excited Kuiper Belt Objects. The common water-ice absorption depths and primordial origins, and the association of Phoebe's water-rich regions with its impact basins, suggests the plausible idea that Kuiper Belt Objects also originated with water-poor surfaces that were enhanced through stochastic collisional modification.
\end{abstract}

\maketitle

\section{Introduction \label{sec:Intro}}

It is generally accepted that the irregular satellite, Phoebe, is a captured body from the outer Solar System. This idea is supported by its low, $\sim8\%$ albedo, and retrograde orbit \citep{Johnson2005}. Orbits such as Phoebe's are a natural bi-product of capture during the early reorganization of the gas-giants which was also responsible for the dispersal of the proto-planetesimal disc \citep{Nesvorny2007}, implying the Phoebe broadly shares the same primordial planetesimal population as the dynamically excited Kuiper Belt Objects \citep[KBOs; ][]{Levison2008}. Admittedly, Phoebe's density of $1.6 \mbox{g cm$^-3$}$ is roughly a factor of two higher than the typical densities of similarly sized KBOs \citep[see for example][]{Grundy2015}.

During its entry into the Saturnian system, the Cassini spacecraft flew past Phoebe, acquiring resolved optical imaging and optical-NIR spectral imaging with the Imaging Sub System (ISS) and Visible-Infrared Mapping Spectrometer (VIMS). ISS imagery revealed a roughly spherical body, with a collisionally evolved surface possessing numerous large impact basins \citep{Porco2005}. The irregular satellite environments around the gas-giants are the most collisionally extreme environments in the known Solar System \citep{Bottke2010}. As a result, it may be that Phoebe has undergone significantly higher bombardment than typically experienced by KBOs. This bombardment may explain Phoebe's comparatively high density.

Rough compositional mapping has been done with the VIMS observations \citep{Clark2005,Hansen2012}. Notable discoveries include the discovery of Fe-rich silicates, CO$_2$ ice, and water absorption \citep{Clark2005, Coradini2008}, with non-negligible absorption of all three found across Phoebe's imaged surface \citep{Cruikshank2010, Hansen2012}. Other candidate materials include C-N organics, consistent with Phoebe's low geometric albedo. Generally however, Phoebe's surface can be simply described as an icy rock. 

Kuiper Belt Objects exhibit a broad range of water-ice absorptions \citep{Barkume2008, Guilbert2009, Brown2012}. No hypothesis has been put forth to explain the origins of this variation. If Phoebe does indeed share a primordial origin with the excited Kuiper Belt Objects, Phoebe's resolved spectral imaging may provide insight into the origin of this water-ice variation between KBOs.

Using an automatic technique to correct the flyby geometry of each VIMS image, we produce the first VIMS maps that are procedurally associated with Phoebe's surface. Using both the optical and IR channels of 42 VIMS images, we produce the most comprehensive compositional maps of Phoebe's surface to date. From this, we infer properties of the water distribution across Phoebe's surface. In particular, we find that Phoebe's primordial surface was likely water poor. The large basin-forming impacts that Phoebe has undergone are responsible for enhancing the surface water content through dredge-up of water-rich sub-surface material. The three spectral units we have identified on Phoebe's surface exhibit water-ice absorptions compatible with the broad range of water-ice absorption exhibited by KBOs \citep{Brown2012}. This implies that large impacts may be the source of the water-ice variations seen from KBO to KBO; the KBOs that exhibit the deepest water-ice absorptions are those that have undergone the most collisional bombardment.

In Section~2 we present our data analysis and mapping techniques. In Section~3 we present water absorption and crater count maps across, along with an analysis of the 3 spectral units we have identified. Finally, in Section~4, we discuss the origin of Phoebe's water distribution, and what that implies for KBOs.

\section{VIMS Data and Processing \label{sec:dataproc}}

We made use of the high and low resolution Visual Image Mapping Spectrometer (VIMS) \citep{Brown2004} observations of Phoebe during the approaching and receding legs of the Phoebe fly by.  A number of image processing and modelling techniques were utilized to produce water absorption maps directly associated with the Phoebe shape. A summary of these steps follows.

\begin{figure}[h]
\includegraphics[width=12cm,angle=270]{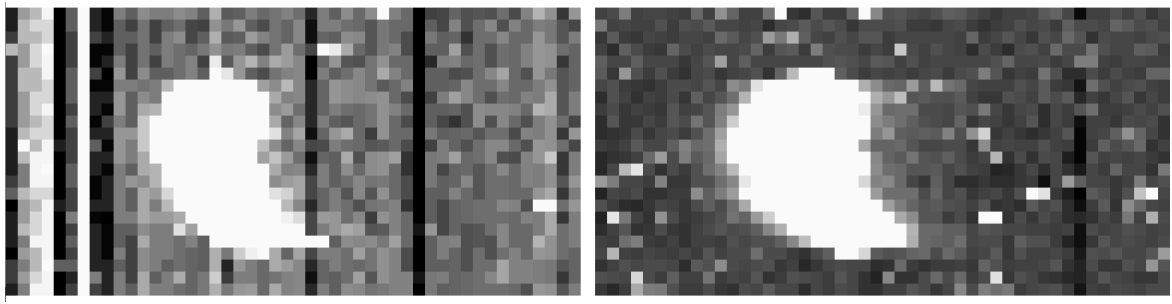}
\caption{Visual channel VIMS image before (left) and after (right) stripe removal. Scaling was set to enhance background levels.\label{fig:striping}}
\end{figure}

VIMS data were first processed through normal ISIS\footnote{\url{https://isis.astrogeology.usgs.gov/index.html}} routines to generate SPICE kernels\footnote{\url{https://naif.jpl.nasa.gov/naif/data.html}}, radiometrically calibrate the raw data (RC17 calibration), and produce VIMS cubes in I/F units in the format of the Flexible Image Transport System\footnote{\url{https://fits.gsfc.nasa.gov/fits_standard.html}}.
Visual channel cubes underwent a custom stripe noise removal routine. The pixels containing only background were used to measure the median value in each image column. That value was removed from each column, and the process was then repeated for each row. This resulted in flat, de-striped and background removed visual channel VIMS images. An example of this de-striping is shown in Figure~\ref{fig:striping}. 

Manual cosmic ray and bad pixel rejection was then performed. Spectra were visibly searched for poorly behaved wavelength regions. These regions were identified by large positive and negative relative variations in the flux at that wavelength between neighbouring pixels in an image. Those wavelengths were ignored. 

To enable modelling of the Cassini-Phoebe geometry, binary images with a pixel value of 1 if that pixel contained Phoebe, 0 if not, were produced for both channels of each VIMS image cube. Modelling of the Phoebe-Cassini geometry two iterations of a forward modelling process, in which a given Sun-Phoebe-Cassini geometry was used to calculate the illumination pattern directly from the medium resolution Gaskell shape model. 

Starting from the observation geometry contained in the SPICE kernel of a cube,  the Sun-Phoebe-Spacecraft geometry was adjusted using a Monte Carlo Markov-chain routine to produce satisfactory matches to both the visual and IR binary images simultaneously. Free parameters in the modelling included the sub-Solar longitude and latitude, the sub-Cassini longitude, latitude, and azimuth, and the Phoebe-Cassini distance at the start of an exposure. An in-image-plane linear relative velocity between Phoebe and Cassini was also fitted to account for tracking errors apparent in most images.  All other parameters, including the Phoebe-Cassini velocity vector were taken as reported in the SPICE kernels. Each pixel from both channels was modelled individually to account for the small changes in the Phoebe-Cassini-Sun geometry that occurred while the image was gathered. For 35 images, a satisfactory geometry was found. 

\begin{figure}[h]
\includegraphics[width=14cm]{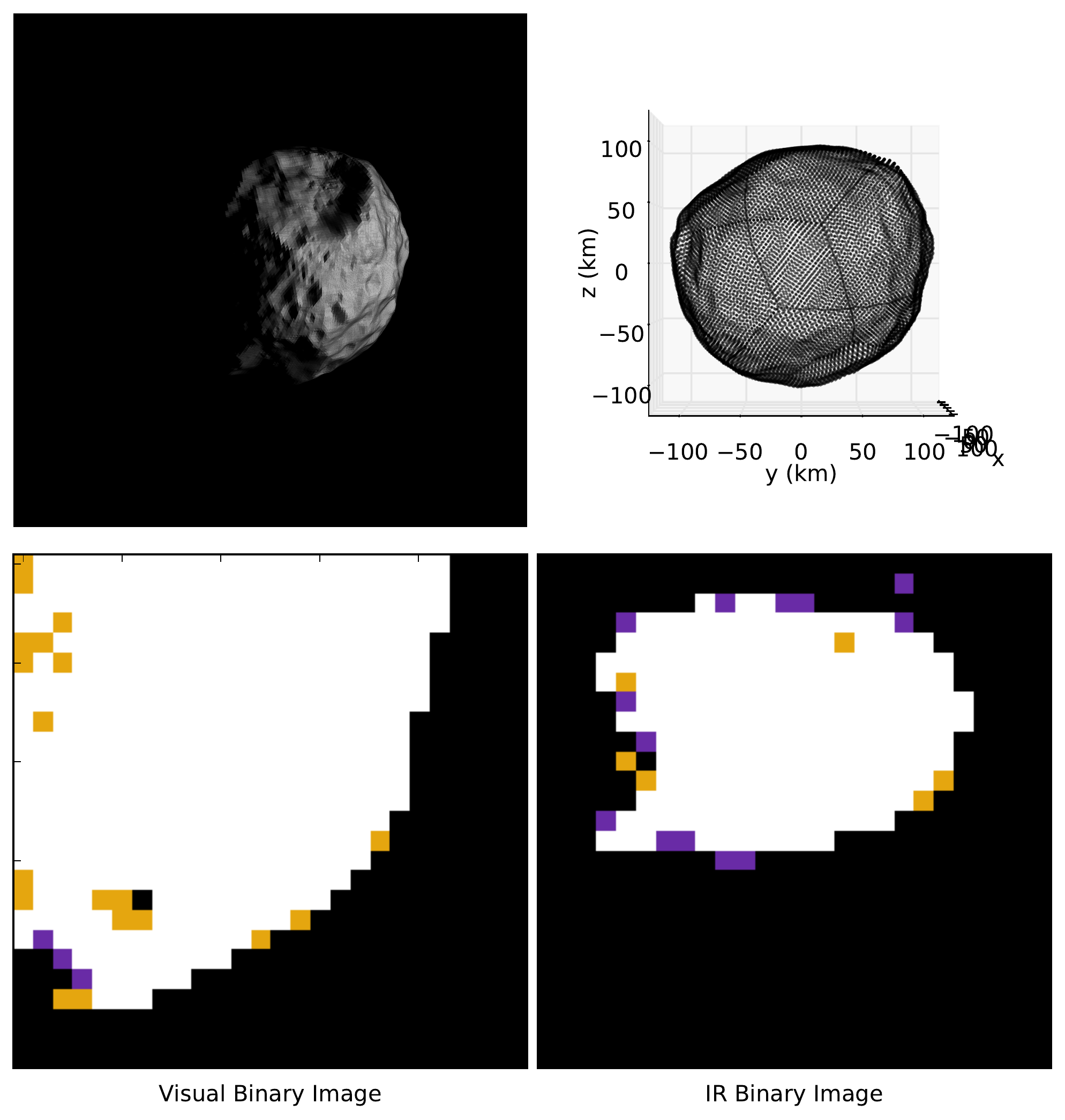}
\caption{Shape model and binary image representations of a successful geometry modelling of cube 1465670650\_1. Surface and shape model renders at the matching geometry are presented in top left and right. Bottom row presents the binary images (1 if a pixel contains an illuminated Phoebe, 0 otherwise) of the VIS (left) and IR (right) channels. Purple are image pixels, orange are model image pixels, and white are where both model and true images contain Phoebe. \label{fig:match}}
\end{figure}

The satisfactorily fitted image geometries were compared to the Cassini-Phoebe ephemeris provided by the Jet Propulsion Laboratory (JPL) Horizons service. Systematic offsets of $1.5^\circ$ and $3.57^\circ$ in the sub-Solar longitude and latitude, $7.71^\circ$ in the sub-Cassini longitude, and $6.7\%$ in the Phoebe-Cassini distance were found. A second stage of geometry matching was applied to all VIMS images with the sub-Solar longitude and latitude, sub-Cassini longitude, and Phoebe-Cassini distances fixed to the adjusted Horizons ephemeris. This resulted in improved matches to the binary images, and enabled 5 additional images to be matched. It was found that systematic offsets of  $-4.05^\circ$ and $8.75^\circ$ applied respectively to the JPL sub-Cassini latitude during the approaching and receding legs of the flyby provided an excellent approximation to the second stage VIMS cube matching results. A third and final match stage was run with the sub-Cassini and sub-Solar points, and the Phoebe-Cassini distance fixed to the offset JPL values, resulting in slight improvements over stage two fits to most image cubes. An example of a successful fit is shown in Figure~\ref{fig:match}.

For each matched VIMS cube, individual spectra were mapped onto the facets of the medium resolution Gaskell shape model for Phoebe. Specifically, separately for each VIMS channel, each \emph{illuminated} facet of the medium resolution shape model was assigned the spectrum associated with the VIMS pixel that imaged those particular facets. For each facet that was imaged 3 or more times, median spectra were produced for each channel. Finally, using the overlap wavelength range, the IR median spectrum of each facet was normalized to that facet's VIS median spectrum, and combined to produce full VIR spectra. for each imaged facet of the shape model.

A list of the 40 images for which a satisfactory geometry match could be found are listed in Table~\ref{tab:geometry}. We group the geometry-matched VIMS cubes into high, medium, and low resolution data groups, with pixel widths at Phoebe of $w<22$, $22<w<50$, and $w>50$~km, respectively. When determining the spectrum of a shape model facet, the high resolution data was utilized if that facet was sufficiently imaged by the high resolution observations. Then, only if a facet did not yet have a median spectrum was the medium resolution data considered, and again for the low resolution data. In this way, the spectral resolution of the global map was as high as possible, but variable. 

Absorption band depths were measured in the usual way, $BD_\lambda = \frac{I/F_\textrm{cont} - I/F_\lambda}{I/F_\textrm{cont}}$ for the 1.6, and 2~$\mu$m water-ice, and 3~$\mu$m ice/hydroxl feature. Linear continua were estimated by interpolating between the fluxes measured at 1.35, 1.78, 2.23 and 3.6~$\mu$m. As the 1.6 and 2~$\mu$m features are associated with pure water-ice, we add the band depths of both together in our subsequent analysis. From these values, simple cylindrical projection absorption maps were produced using each facet's mean latitude and longitude with respect to the shape model centre. We chose to forgo any attempts to extract specific water abundances through spectral modelling as determining correct abundances depends on knowing the exact chemical and physical properties of Phoebe's surface. Rather, we restrict our discussion purely to absorption depths.

Albedos were assigned to each facet of the shape model using the comparatively low resolution albedo map produce from visible Voyager imaging of the satellite \citep{Simonelli1999}. No higher resolution albedo map had been produced utilizing the Cassini flyby data. The mean latitude and longitude of each shape model facet was calculated with respect to the shape model origin, and an albedo for that facet was sampled from that latitude and longitude on the Voyager albedo map.

Craters were counted in a very simple fashion, using the high resolution surface map\footnote{\url{http://www.ciclops.org/view/1743/Map-of-Phoebe---December-2005}} produced by the Cassini Imaging Team. Craters were manually identified and their radii measured directly off the map. Only crater-like features with radii at least 3 pixels (798 metres on Phoebe's surface) were counted. While the count was performed across the entire map, the smallest craters were only detectable in the highest resolution regions (longitudes, $0 \leq L \leq 90^\circ$). To avoid significant uncertainty in crater radius, we limited our analysis of the crater size frequency distribution to those craters with radii larger than 5 image pixels, or 1.15~km at zero degrees latitude. A more robust and complicated approach would be to count directly from the Imaging Subs System \citep[ISS;][]{Porco2004} images after a deprojection, rather than the global image mosaic. That more complicated approach, while not affecting the identification of large craters, would enable more reliable detection and more accurate radius measurement of the small, $r\lesssim0.8$~km, craters. This extra level of complication would not however, afford a different interpretation to be made regarding the variation in the density of craters with radii $1\lesssim r\lesssim3$~km between water rich and poor regions. We opt to forgo this unnecessary level of complication.

To understand the distribution of water in Phoebe, we consider a simple global shape from which depth can be measured.  Phoebe's shape is nearly ellipsoidal, with only the impact basins showing deviations away from ellipsoidal shape of more than a few kilometres. To demonstrate this, a triaxial ellipsoid was fit to the Phoebe shape model. The ellipsoid shape, rotation, and origin were allowed to vary until the root-mean-square (RMS) residual value between the Gaskell shape model and the ellipsoid was at a minimum. The best-fit ellipsoid had an RMS of 3.0~km, and had shape parameters, $a = 107.7$~km, $b= 107.7$~km, and $c=104.1$~km. This fit avoided the Jason and South Pole impact basins, though it should be noted that their inclusion in the fit had almost no affect at all, resulting in a nearly identical shape ($a=107.1$~km, $b=107.0$~km, and $c=105.0$~km) and rotation of the ellipsoid, though with a larger RMS of 4.1~km. Depth of each shape model facet was evaluated as the distance between the facet and the best-fit ellipsoid along the line that connected the ellipsoid origin and the facet itself. Virtually identical depths were found when using either fitted ellipsoid, or even a sphere of radius equal to the mean radius of Phoebe (106.6~km).

For ease of presentation, we emphasize multiple distinct regions covered by the high resolution data group: the Jason impact basin, the water-poor regions east and west of Jason which occupy a similar latitude range, and the water-rich region south of the Jason impact basin. We also point out the south pole impact basin which is predominantly covered by the medium and low resolution data, a region where water absorption appears strongest.

%In simulating HST colours, we pass through synphot, the Kurcurz Solar spectrum model convolved with the spectra.
\section{Mapping Results\label{sec:mapping}}

In the following sections, we present our cylindrical absorption depth, albedo, and crater density maps. We discuss each of these maps in turn.

\subsection{Spectra and Spectral Maps}
\begin{figure}[h]
   \epsscale{1.2}
   \centering
   \plotone{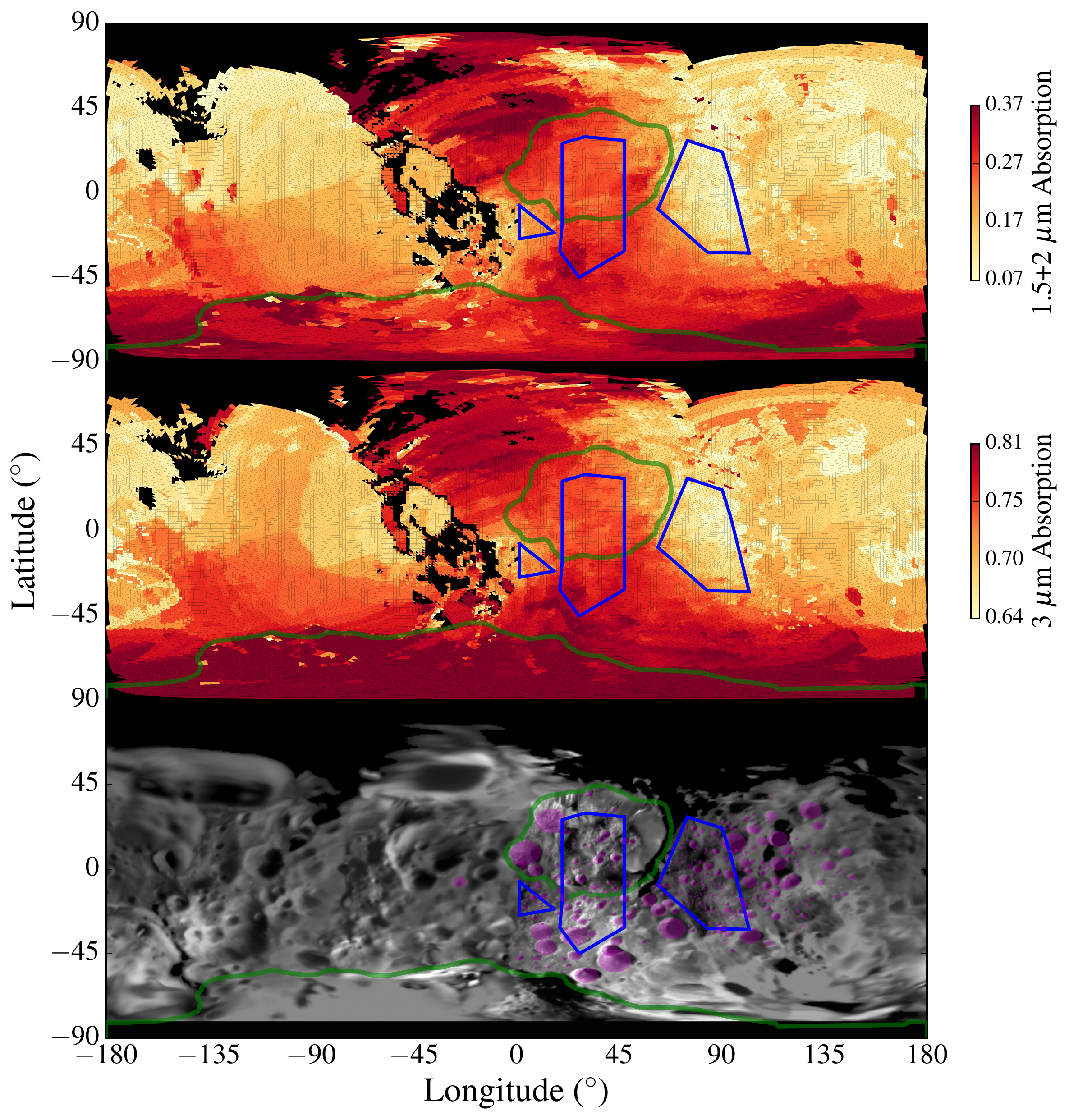} 
   \figcaption{Water absorption band depth, and optical image maps. 1.6+2~$\mu$m absorption depth is shown at top, and the 3~$\mu$m absorption depth shown in middle. Traced on the bottom map are the two basin edges shown in green. The locations of all craters counted in the high resolution data are marked by the purple circles. The two water-poor (left and right) and water-rich (middle) regions from which high resolution crater radius functions are generated are outlined in blue. Note the high latitudinal coverage of these maps are a result of illumination of Phoebe's non-spherical Northern region, and not in contradiction with the sub-Solar latitude of $-12.2^\circ$ during the flyby. \ \label{fig:fullMap}}
\end{figure}

\begin{figure}[h]
   \centering
   \includegraphics[width=\textwidth,angle=270]{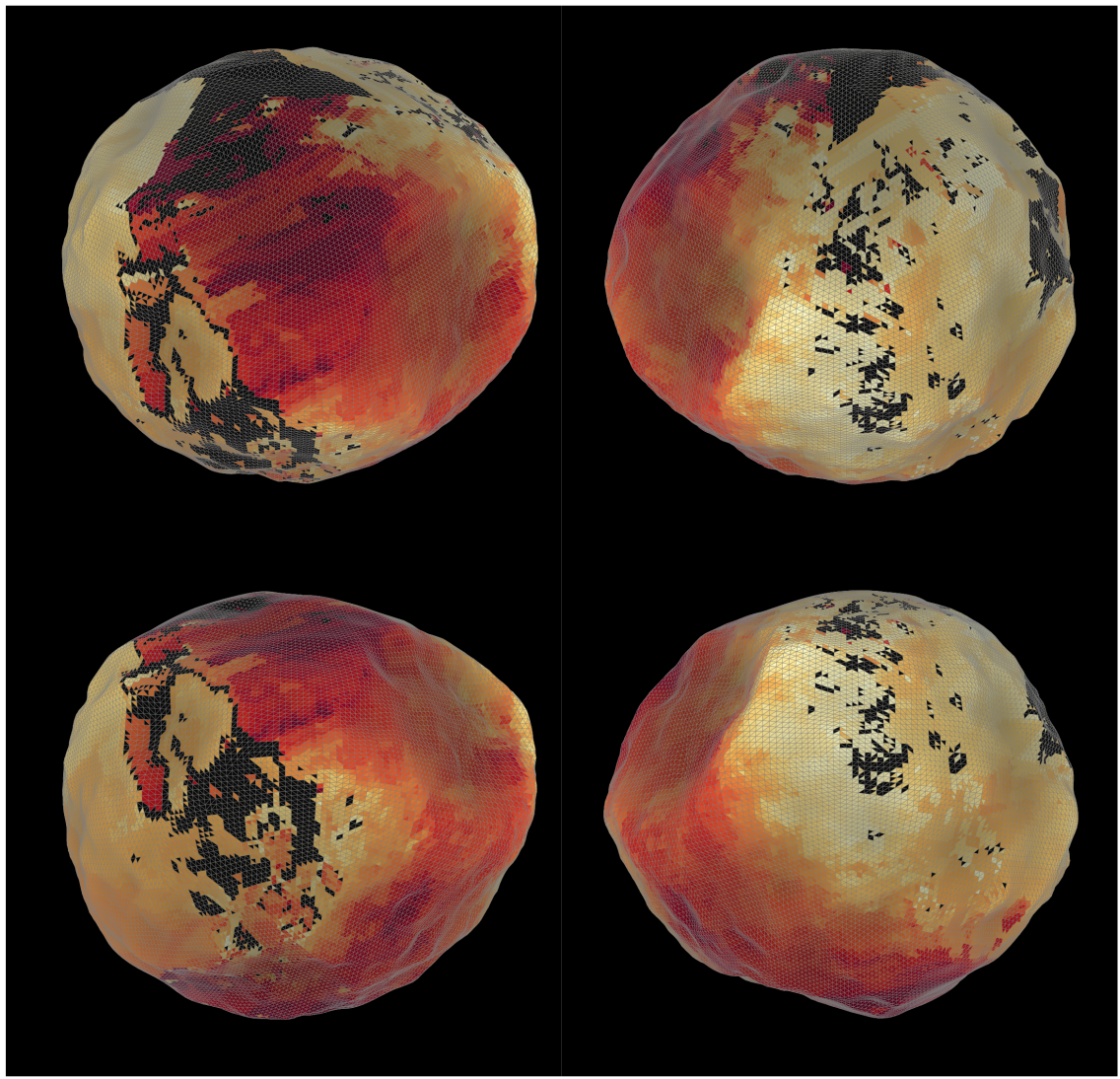} 
   \figcaption{Water-ice absorption projected onto the medium resolution shape model. The colour scale is the same as that used in Figure~\ref{fig:fullMap}. The sub-viewer longitudes are $-15^\circ$ (left) and $87^\circ$ (right) and latitudes are $35^\circ$ (top) and $-10^\circ$ (bottom) and were chosen to show all faces of Jason Crater. \label{fig:shape}}
\end{figure}

Full water absorption depth maps are presented in Figure~\ref{fig:fullMap}. The colour scale was chosen to emphasize structure in the maps. In Figure~\ref{fig:shape} we present the water-ice absorption colour map projected onto the Phoebe shape model.

Generally, we confirm the broad findings of \citet{Clark2005} in that water is found everywhere on Phoebe; there is no region with spectra completely devoid of the 1.6, 2, and 3~$\mu$m features. Our maps roughly produce a similar distribution of water across Phoebe's surface as those of \citet{Hansen2012}. Though with the broader coverage of our maps, additional details can be seen. The regions richest in water are clearly associated with the impact basins, with the iciest areas found just beyond the outer edges of the Jason and South Pole basins. The floor of the South Pole basin exhibits significantly deeper 3~$\mu$m absorption than found inside Jason Basin. The majority of water is concentrated in a $\sim 70^\circ$ wide nearly latitudinal strip encompassing both impact basins (Figure~\ref{fig:shape}) 

\begin{figure}[h]
   \epsscale{1.2}
   \centering
   \plotone{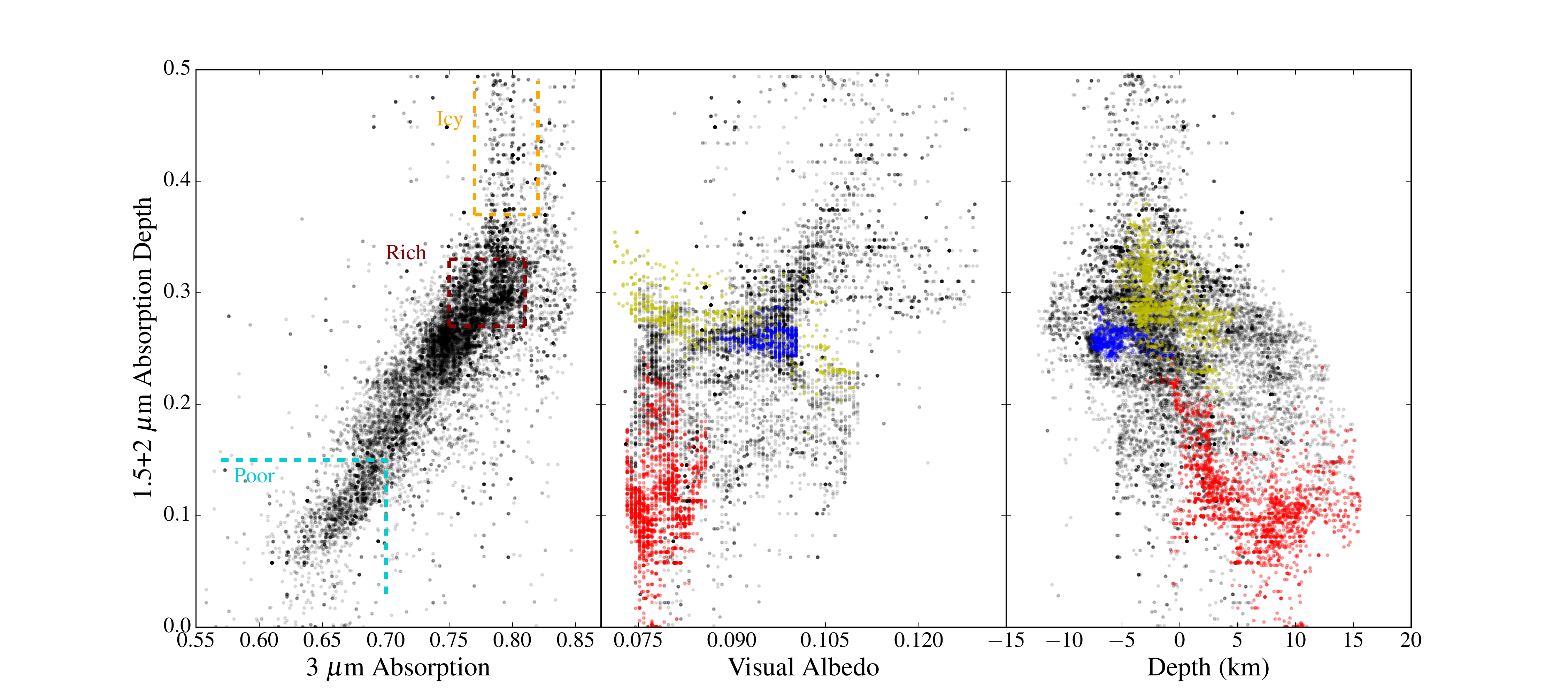} 
   \figcaption{Band depth vs. band depth, albedo and elevation. The regions in which average spectra were generated for the poor, rich, and icy spectral types are plotted. In the middle and left plots, blue, yellow, and red points are sampled from inside Jason Crater, the water rich region South of Jason, and the water-poor region to the east of Jason, respectively. Only the points from facets covered by the high resolution VIMS datagroup are presented. \label{fig:correlations}}
\end{figure}

We present the 1.6+2~$\mu$m band depth versus the 3~$\mu$m feature band depth, and albedo in Figure~\ref{fig:correlations}. Unlike the findings of Clark et al. (2005) who found Phoebe to exhibit increasing water content with depth, we find the opposite. The majority of Phoebe's surface exhibits a clear trend of increasing water absorption with physical depth, both locally, and globally. While physical depth cannot be mapped to water absorption uniquely, each distinct region generally exhibits the same positive correlation between water absorption and physical depth. The only exception is Jason crater, which exhibits water rich walls, and a less rich crater floor (see Figure~\ref{fig:shape}). This aspect of Jason crater led Clark to conclude that Phoebe's surface layers are richer in water than the sub-surface layers. For any other region, and globally, the opposite is certainly true. 

\begin{figure}[h]
   \centering
   \plotone{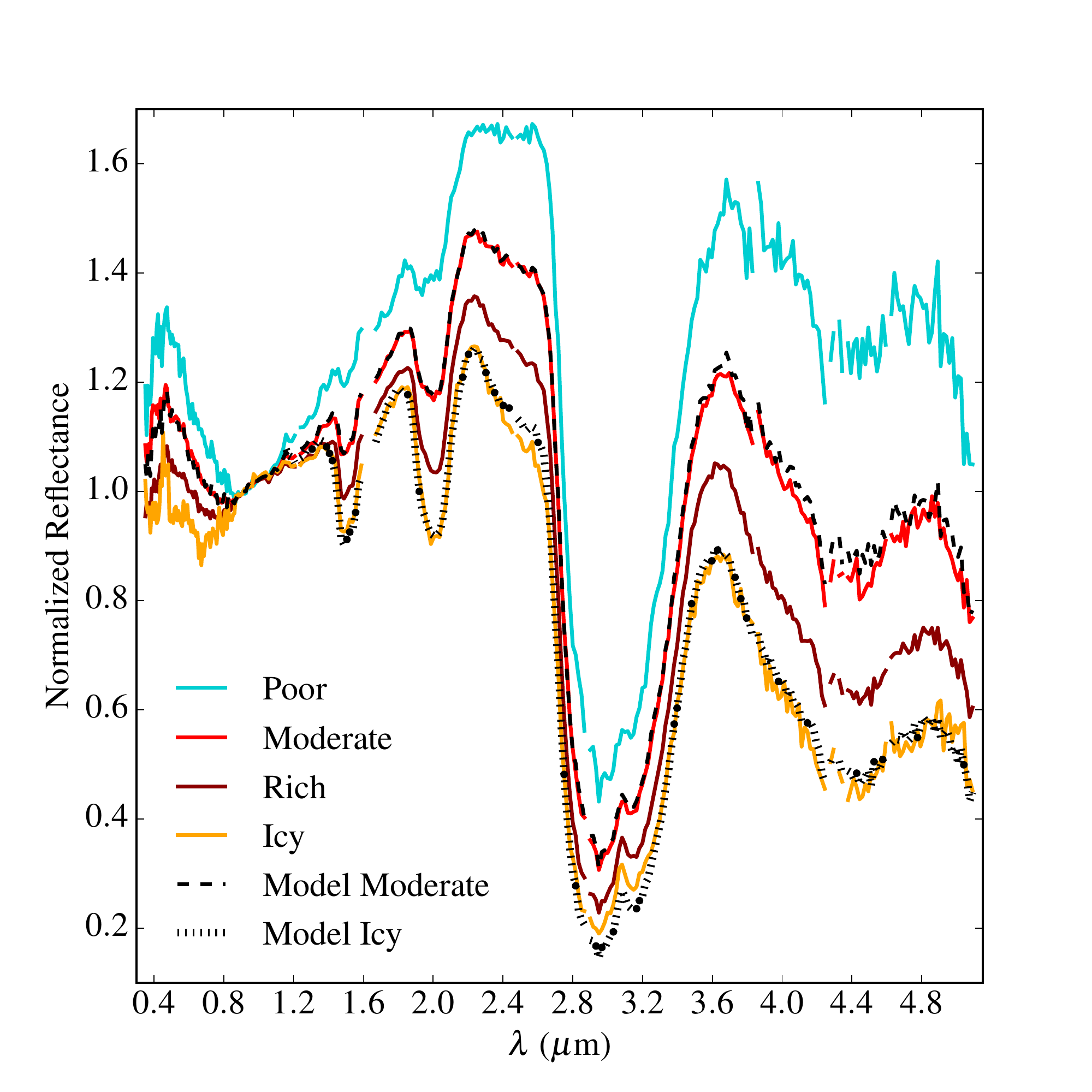} 
   \figcaption{Spectra of the different spectral types. Modelled spectra of the Moderate and Icy spectral types are shown in the dashed and hashed lines respectively. \label{fig:spectra}}
\end{figure}

Phoebe's surface exhibits positively correlated water-ice and ice/hydroxl band depths. From water content considerations, Phoebe exhibits only three spectral types. The majority of the surface falls along a tight linear continuum with water-rich and water-poor end members (see Figure~\ref{fig:correlations}). The average end member spectrum, including an intermediate spectrum are presented in Figure~\ref{fig:spectra}. As shown in that figure, the linear trend of ice and hydroxl absorption depth, and indeed the vast majority of spectral behaviour along that trend can be very simply accounted for by a geographic mixture of the two end members. Only a small fraction of Phoebe's surface is of a third type, and is not a product of a mixture of the other two. Rather, these icy regions posses deeper 1.6 and 2~$\mu$m absorption depths than even the water-rich regions. These spots, predominantly found on shiny basin walls, are particularly icy compared to their surroundings. The spectra of these regions can broadly be accounted for by a geographic mix of the rich surface end member and pure crystalline water-ice (see Figure~\ref{fig:spectra}).

As shown in Figure~\ref{fig:correlations}, visual albedo generally correlates positively with water-ice absorption depth. This implies that the presence of water, in ice form, largely governs the visual albedo. We caution against drawing further conclusions from the specific structures seen in the water-ice versus albedo plot, as the Voyager albedo map is of a poor resolution compared to the absorption maps. It is unclear how much of the plot's fine structure is real.

\subsection{Crater Counts}
\begin{figure}[h]
   \centering
   \plotone{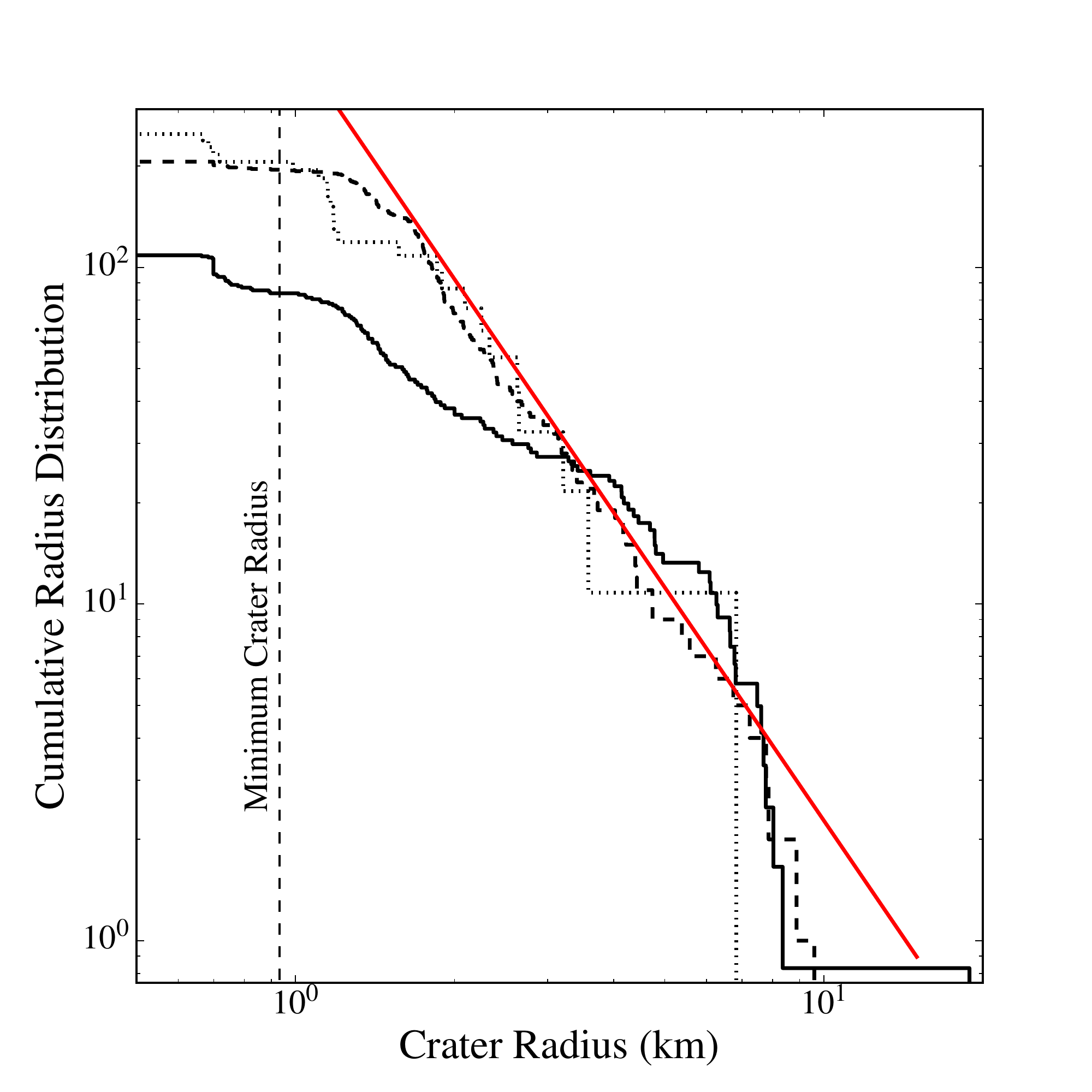} 
   \figcaption{Cumulative crater distributions. Dashed and dotted lines demark the crater counts in the water-poor regions East and West of Jason, while the solid line demarks the counts and the water-rich regions inside and to the South of Jason. The distributions have been scaled  by the ratio of their areas and the area of the larger water-poor region. The red line has slope $q=-2.3$. \label{fig:craters}}
\end{figure}

While we counted craters across the entire surface map, we limit our discussion to the regions with the highest ISS image resolution; Jason basin and south of Jason basin, and in the water-poor regions to the East and West. These regions and counted craters are marked in Figure~\ref{fig:fullMap}. We present crater cumulative radius distributions (CRD) in Figure~\ref{fig:craters}. The CRDs we produce are broadly compatible with past crater counting efforts \citep{Porco2005}. No significant variations in the water-rich crater CRDs inside or south of Jason basin were found. The water rich and poor regions exhibit dramatically different crater CRDs. While water-rich and poor regions exhibit a nearly equal density of craters with radii $r\gtrsim5$~km, the water-poor regions are much richer in smaller craters possessing nearly 2.5 times more craters with $r\sim1$~km than the water-rich regions. Furthermore, the water-poor CRD is nearly collisional, being well represented by a power-law distribution $N(>r)\propto r^{-q}$ with slope of $q\sim-2.3$. The water-rich CRD exhibits an inflection at $r\sim5$~km from a steeper than collisional slope to a shallower than collisional distribution with slope $q\sim-1$. For craters larger than our completion limit, the Kolmogorov-Smirnoff test suggests that the probability that the area-normalized CRDs of the water-rich and poor regions are drawn from the same parent distribution is only 0.006\%. It is clear even from our simple crater counting method that the water-poor regions have significantly higher small crater density than do the water-rich regions. 

Beyond simple crater counts, visibly the largest craters differ in profile between the water-rich and poor regions. While slumping is apparent in the largest craters across the entirety of Phoebe, the crater size at which crater wall slumping is apparent is significantly smaller in the water-rich region. That is, moderate sized craters appear much more conical in shape than similar sized craters in the water poor region.

It is only reasonable that the impactor distribution on Phoebe is broadly homogenous, with no preference for small objects to impact either water-rich or poor regions. It seems likely therefore, that both terrain types share roughly the same crater formation distributions. The difference in crater CRDs and the difference in shapes of larger craters appear to be the the result of water content, resulting in different rheologies for the two terrain types. It seems that water-ice acts as a lubricant, allowing slumping and small crater hiding - possibly driven through impact induced Phoebe quakes - to occur more readily.

\section{Origin of Phoebe's Water Distribution \label{sec:discussion}}

The clear concentration of water in and around Phoebe's two large impact basins argues that the impacts which formed those basins have modified the water distribution, enhancing water absorption on Phoebe's surface. The positive correlation of depth and water absorption, suggests a plausible scenario that the early Phoebe possessed a water-poor surface, with richer subsurface layers. The two large basin forming impacts then exposed those deeper, water rich layers, and enhanced the water content of the regions surrounding those basins. 

Surface water-ice is stable at Saturn's distance from the Sun. As has been suggested from dynamical considerations, Phoebe plausibly originated from a reservoir of planetesimals that was beyond Saturn's current orbit \citep{Nesvorny2007} and therefore, is unlikely to have spent long durations with significantly warmer surface temperatures than it has experienced since capture by Saturn. It would follow that Phoebe's originally low surface water concentration was not from post-formation heating, but rather, a property it possessed while residing in the planetesimal reservoir from which it came. It seems reasonable to expect that the majority of objects such as the Jupiter Trojans, and Kuiper belt Objects (KBOs) which originated from the same population as Phoebe also originally possessed equally water-poor surfaces.

\begin{figure}[h]
   \centering
   \plotone{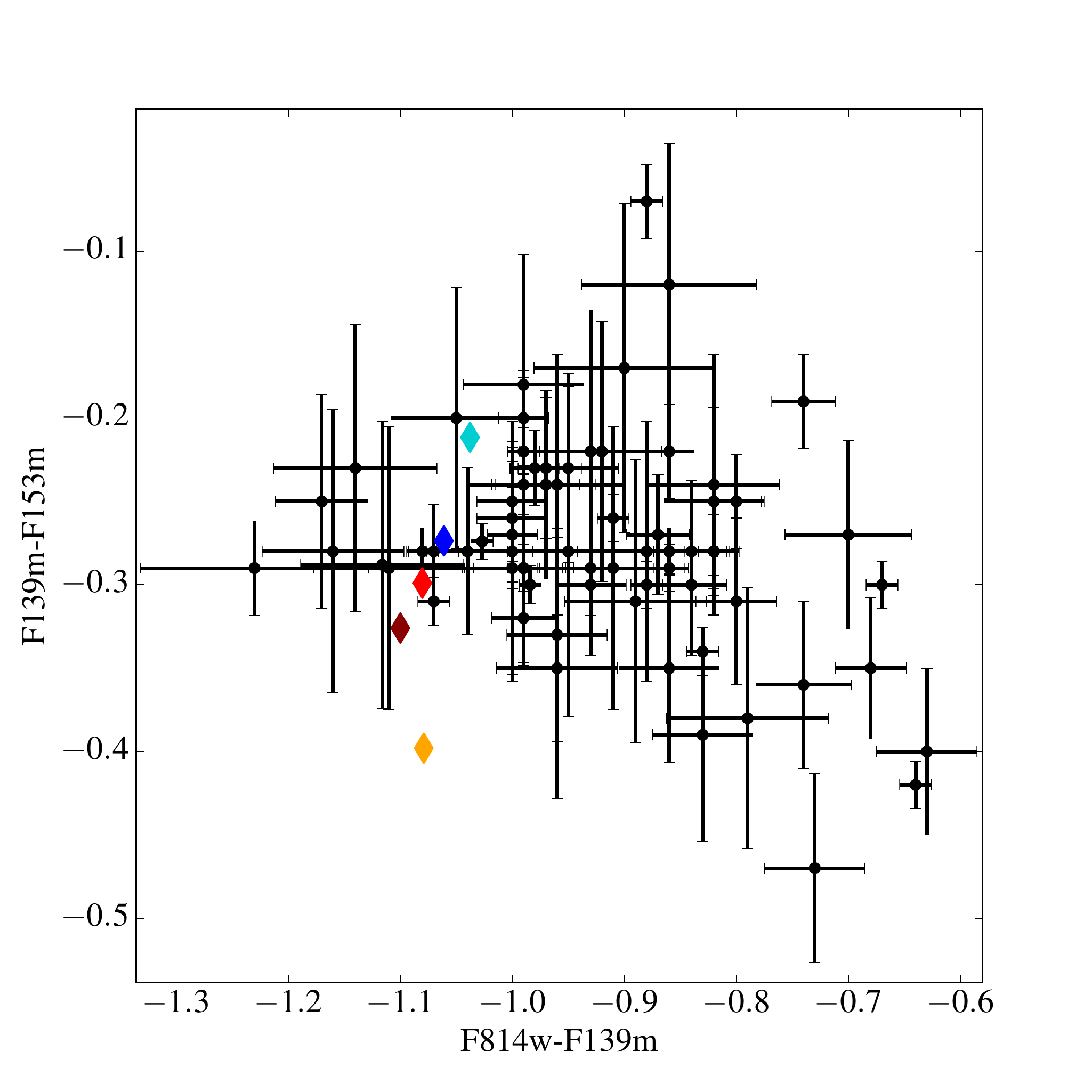} 
   \figcaption{Observed HST NIR and water-ice absorption colours of small KBOs and simulated Phoebe colours (diamonds). The simulated points have been coloured to match the spectra presented in Figure~\ref{fig:spectra}. The colour of the average Phoebe spectrum is shown by the red diamond. \label{fig:HSTcolours}}
\end{figure}

Between the epoch of formation and their scattering into the outer Solar System, KBOs are unlikely to have experienced hotter temperatures than Phoebe's current surface temperature. Therefore, like Phoebe, sublimation processes are unlikely to have affected the surface water concentrations on KBOs. The broad range of water-ice absorptions exhibited by KBOs \citep[see for example][]{Barkume2008,Brown2012} is broadly compatible with the variation in absorptions across Phoebe's surface. To compare the range of water-ice absorptions exhibited by Phoebe's different spectral types, and that of KBOs, we estimated the colours of those spectral types in the Hubble Space Telescope filter system using the \emph{synphot} routine of the \emph{stsdas} \emph{IRAF} package. In Figure~\ref{fig:HSTcolours} we plot the (F814w-F139m) colour versus the (F139m-F153m) colour. The latter pf the two colours is sensitive to the $1.5 \mbox{ $\mu$m}$ water-ice absorption feature, while the former provides a measure of the NIR continuum colour. Alongside the colours of Phoebe's spectral types, we plot the observed colours of a large sample of KBOs in Figure~\ref{fig:HSTcolours} \citep{Fraser2015}.

Phoebe's surface exhibits the same range (F139m-F153m) colour as do small KBOs. In particular, the colour of the water-rich and water-poor spectral types match the colours of the bluest and reddest KBOs, respectively. Broadly, this implies that KBOs exhibit a range of water-ice concentrations similar to that observed on Phoebe. If Phoebe and the dynamically excited KBOs share the same primordial origins, it seems plausible that large basin forming impacts could be responsible for enhancing the water-ice concentrations on KBOs like has occurred on Phoebe. The stochastic nature of those impacts is compatible with the observation that not all KBOs exhibit equal levels of water-ice absorption. This idea would imply that KBOs with the highest water-ice concentrations are those that experienced the highest levels of collisional dredge up. KBOs with the lowest water-ice concentrations possess old surfaces that have undergone little to no collisional dredge up. An extreme version of this idea is perpetuated by the Haumea collisional family, the members of which are purported to be ejecta fragments of the large KBO Haumea \citep{Brown2007}. Those fragments, and Haumea itself exhibit the deepest water-ice absorptions of any KBO, implying a deep dredge-up origin for those objects. Other observed KBOs have not undergone equally catastrophic collisions capable of the dramatic re-surfacing Haumea experienced.

A notable difference exists between Phoebe, and other small KBOs; other than the Haumea family members, small KBOs exhibit optical and NIR continuum colours which anti-correlate with the (F139m-F153m) colour \citep{Fraser2012}. Put another way, within a KBO compositional class, the redder a KBO the more water-ice absorption it exhibits \citep{Brown2012}. Phoebe does not show this behaviour, exhibiting no appreciable change in NIR continuum colour, (F814w-F139m) for any of its spectral types. Moreover, Phoebe exhibits significantly bluer optical colours than do KBOs. It is hypothesized that KBO's posses by organic materials that are associated with the water-ice components on those bodies \citet{Fraser2012}. This results not only in their red colours, but also the correlation the extremity of a KBOs redness and its water ice absorption depth. What ever organics are responsible for the red colours of KBOs, its clear this material does not exist in any appreciable quantities on Phoebe's surface. The heavy collisional bombardment Phoebe has experienced in the Saturnian environment may be responsible for the removal of this material. It has been observed that some centaurs lose their red colouring, and transition to purely neutral surfaces at a similar point to when they dynamically transition from the centaur region to the Jupiter Family comet region \citep{Jewitt2015}, a similar heliocentric distance to Saturn. It may be then that what ever thermal processes cause centaurs to lose their colouring could also have acted on Phoebe, resulting in its neutral surface colour.

\acknowledgements

The authors would like to thank Pedro Lacerda for his useful insights during discussions of this work. This research has made use of the USGS Integrated Software for Imagers and Spectrometers (ISIS). Raw data were generated at the Planetary Data System Cassini Archive (\url{http://pds-atmospheres.nmsu.edu/data_and_services/atmospheres_data/Cassini/Cassini.html}). Processed data products are available from the corresponding author upon request (wes.fraser@qub.ac.uk). W.C.F. acknowledges support from Science and Technologies Funding Council grant ST/P0003094/1. M.E.B. and W.C.F. acknowledge support from the National Aeronautics and Space Administration Cassini Data Analysis program grant NNX10AF11G.

\software{USGS Integrated Software for Imagers and Spectrometers}, \software{Image Reduction and Analysis Facility}, \software{Space Telescope Science Data Analysis System}

\bibliographystyle{apj}
\bibliography{phoebe} 

\begin{thebibliography}{20}
\expandafter\ifx\csname natexlab\endcsname\relax\def\natexlab#1{#1}\fi

\bibitem[{{Barkume} {et~al.}(2008){Barkume}, {Brown}, \&
  {Schaller}}]{Barkume2008}
{Barkume}, K.~M., {Brown}, M.~E., \& {Schaller}, E.~L. 2008, \aj, 135, 55

\bibitem[{{Bottke} {et~al.}(2010){Bottke}, {Nesvorn{\'y}}, {Vokrouhlick{\'y}},
  \& {Morbidelli}}]{Bottke2010}
{Bottke}, W.~F., {Nesvorn{\'y}}, D., {Vokrouhlick{\'y}}, D., \& {Morbidelli},
  A. 2010, \aj, 139, 994

\bibitem[{{Brown} {et~al.}(2007){Brown}, {Barkume}, {Ragozzine}, \&
  {Schaller}}]{Brown2007}
{Brown}, M.~E., {Barkume}, K.~M., {Ragozzine}, D., \& {Schaller}, E.~L. 2007,
  \nat, 446, 294

\bibitem[{{Brown} {et~al.}(2012){Brown}, {Schaller}, \& {Fraser}}]{Brown2012}
{Brown}, M.~E., {Schaller}, E.~L., \& {Fraser}, W.~C. 2012, \aj, 143, 146

\bibitem[{{Brown} {et~al.}(2004){Brown}, {Baines}, {Bellucci}, {Bibring},
  {Buratti}, {Capaccioni}, {Cerroni}, {Clark}, {Coradini}, {Cruikshank},
  {Drossart}, {Formisano}, {Jaumann}, {Langevin}, {Matson}, {McCord},
  {Mennella}, {Miller}, {Nelson}, {Nicholson}, {Sicardy}, \&
  {Sotin}}]{Brown2004}
{Brown}, R.~H., {Baines}, K.~H., {Bellucci}, G., {Bibring}, J.-P., {Buratti},
  B.~J., {Capaccioni}, F., {Cerroni}, P., {Clark}, R.~N., {Coradini}, A.,
  {Cruikshank}, D.~P., {Drossart}, P., {Formisano}, V., {Jaumann}, R.,
  {Langevin}, Y., {Matson}, D.~L., {McCord}, T.~B., {Mennella}, V., {Miller},
  E., {Nelson}, R.~M., {Nicholson}, P.~D., {Sicardy}, B., \& {Sotin}, C. 2004,
  \ssr, 115, 111

\bibitem[{{Clark} {et~al.}(2005){Clark}, {Brown}, {Jaumann}, {Cruikshank},
  {Nelson}, {Buratti}, {McCord}, {Lunine}, {Baines}, {Bellucci}, {Bibring},
  {Capaccioni}, {Cerroni}, {Coradini}, {Formisano}, {Langevin}, {Matson},
  {Mennella}, {Nicholson}, {Sicardy}, {Sotin}, {Hoefen}, {Curchin}, {Hansen},
  {Hibbitts}, \& {Matz}}]{Clark2005}
{Clark}, R.~N., {Brown}, R.~H., {Jaumann}, R., {Cruikshank}, D.~P., {Nelson},
  R.~M., {Buratti}, B.~J., {McCord}, T.~B., {Lunine}, J., {Baines}, K.~H.,
  {Bellucci}, G., {Bibring}, J.-P., {Capaccioni}, F., {Cerroni}, P.,
  {Coradini}, A., {Formisano}, V., {Langevin}, Y., {Matson}, D.~L., {Mennella},
  V., {Nicholson}, P.~D., {Sicardy}, B., {Sotin}, C., {Hoefen}, T.~M.,
  {Curchin}, J.~M., {Hansen}, G., {Hibbitts}, K., \& {Matz}, K.-D. 2005, \nat,
  435, 66

\bibitem[{{Coradini} {et~al.}(2008){Coradini}, {Tosi}, {Gavrishin},
  {Capaccioni}, {Cerroni}, {Filacchione}, {Adriani}, {Brown}, {Bellucci},
  {Formisano}, {D'Aversa}, {Lunine}, {Baines}, {Bibring}, {Buratti}, {Clark},
  {Cruikshank}, {Combes}, {Drossart}, {Jaumann}, {Langevin}, {Matson},
  {McCord}, {Mennella}, {Nelson}, {Nicholson}, {Sicardy}, {Sotin}, {Hedman},
  {Hansen}, {Hibbitts}, {Showalter}, {Griffith}, \&
  {Strazzulla}}]{Coradini2008}
{Coradini}, A., {Tosi}, F., {Gavrishin}, A.~I., {Capaccioni}, F., {Cerroni},
  P., {Filacchione}, G., {Adriani}, A., {Brown}, R.~H., {Bellucci}, G.,
  {Formisano}, V., {D'Aversa}, E., {Lunine}, J.~I., {Baines}, K.~H., {Bibring},
  J.-P., {Buratti}, B.~J., {Clark}, R.~N., {Cruikshank}, D.~P., {Combes}, M.,
  {Drossart}, P., {Jaumann}, R., {Langevin}, Y., {Matson}, D.~L., {McCord},
  T.~B., {Mennella}, V., {Nelson}, R.~M., {Nicholson}, P.~D., {Sicardy}, B.,
  {Sotin}, C., {Hedman}, M.~M., {Hansen}, G.~B., {Hibbitts}, C.~A.,
  {Showalter}, M., {Griffith}, C., \& {Strazzulla}, G. 2008, \icarus, 193, 233

\bibitem[{{Cruikshank} {et~al.}(2010){Cruikshank}, {Meyer}, {Brown}, {Clark},
  {Jaumann}, {Stephan}, {Hibbitts}, {Sandford}, {Mastrapa}, {Filacchione},
  {Dalle Ore}, {Nicholson}, {Buratti}, {McCord}, {Nelson}, {Dalton}, {Baines},
  \& {Matson}}]{Cruikshank2010}
{Cruikshank}, D.~P., {Meyer}, A.~W., {Brown}, R.~H., {Clark}, R.~N., {Jaumann},
  R., {Stephan}, K., {Hibbitts}, C.~A., {Sandford}, S.~A., {Mastrapa},
  R.~M.~E., {Filacchione}, G., {Dalle Ore}, C.~M., {Nicholson}, P.~D.,
  {Buratti}, B.~J., {McCord}, T.~B., {Nelson}, R.~M., {Dalton}, J.~B.,
  {Baines}, K.~H., \& {Matson}, D.~L. 2010, \icarus, 206, 561

\bibitem[{{Fraser} \& {Brown}(2012)}]{Fraser2012}
{Fraser}, W.~C. \& {Brown}, M.~E. 2012, \apj, 749, 33

\bibitem[{{Fraser} {et~al.}(2015){Fraser}, {Brown}, \& {Glass}}]{Fraser2015}
{Fraser}, W.~C., {Brown}, M.~E., \& {Glass}, F. 2015, \apj, 804, 31

\bibitem[{{Grundy} {et~al.}(2015){Grundy}, {Porter}, {Benecchi}, {Roe}, {Noll},
  {Trujillo}, {Thirouin}, {Stansberry}, {Barker}, \& {Levison}}]{Grundy2015}
{Grundy}, W.~M., {Porter}, S.~B., {Benecchi}, S.~D., {Roe}, H.~G., {Noll},
  K.~S., {Trujillo}, C.~A., {Thirouin}, A., {Stansberry}, J.~A., {Barker}, E.,
  \& {Levison}, H.~F. 2015, \icarus, 257, 130

\bibitem[{{Guilbert} {et~al.}(2009){Guilbert}, {Alvarez-Candal}, {Merlin},
  {Barucci}, {Dumas}, {de Bergh}, \& {Delsanti}}]{Guilbert2009}
{Guilbert}, A., {Alvarez-Candal}, A., {Merlin}, F., {Barucci}, M.~A., {Dumas},
  C., {de Bergh}, C., \& {Delsanti}, A. 2009, \icarus, 201, 272

\bibitem[{{Hansen} {et~al.}(2012){Hansen}, {Hollenbeck}, {Stephan}, {Apple}, \&
  {Shin-White}}]{Hansen2012}
{Hansen}, G.~B., {Hollenbeck}, E.~C., {Stephan}, K., {Apple}, S.~K., \&
  {Shin-White}, E.-J.~Z. 2012, \icarus, 220, 331

\bibitem[{{Jewitt}(2015)}]{Jewitt2015}
{Jewitt}, D. 2015, \aj, 150, 201

\bibitem[{{Johnson} \& {Lunine}(2005)}]{Johnson2005}
{Johnson}, T.~V. \& {Lunine}, J.~I. 2005, \nat, 435, 69

\bibitem[{{Levison} {et~al.}(2008){Levison}, {Morbidelli}, {Van Laerhoven},
  {Gomes}, \& {Tsiganis}}]{Levison2008}
{Levison}, H.~F., {Morbidelli}, A., {Van Laerhoven}, C., {Gomes}, R., \&
  {Tsiganis}, K. 2008, \icarus, 196, 258

\bibitem[{{Nesvorn{\'y}} {et~al.}(2007){Nesvorn{\'y}}, {Vokrouhlick{\'y}}, \&
  {Morbidelli}}]{Nesvorny2007}
{Nesvorn{\'y}}, D., {Vokrouhlick{\'y}}, D., \& {Morbidelli}, A. 2007, \aj, 133,
  1962

\bibitem[{{Porco} {et~al.}(2005){Porco}, {Baker}, {Barbara}, {Beurle},
  {Brahic}, {Burns}, {Charnoz}, {Cooper}, {Dawson}, {Del Genio}, {Denk},
  {Dones}, {Dyudina}, {Evans}, {Giese}, {Grazier}, {Helfenstein}, {Ingersoll},
  {Jacobson}, {Johnson}, {McEwen}, {Murray}, {Neukum}, {Owen}, {Perry},
  {Roatsch}, {Spitale}, {Squyres}, {Thomas}, {Tiscareno}, {Turtle}, {Vasavada},
  {Veverka}, {Wagner}, \& {West}}]{Porco2005}
{Porco}, C.~C., {Baker}, E., {Barbara}, J., {Beurle}, K., {Brahic}, A.,
  {Burns}, J.~A., {Charnoz}, S., {Cooper}, N., {Dawson}, D.~D., {Del Genio},
  A.~D., {Denk}, T., {Dones}, L., {Dyudina}, U., {Evans}, M.~W., {Giese}, B.,
  {Grazier}, K., {Helfenstein}, P., {Ingersoll}, A.~P., {Jacobson}, R.~A.,
  {Johnson}, T.~V., {McEwen}, A., {Murray}, C.~D., {Neukum}, G., {Owen}, W.~M.,
  {Perry}, J., {Roatsch}, T., {Spitale}, J., {Squyres}, S., {Thomas}, P.~C.,
  {Tiscareno}, M., {Turtle}, E., {Vasavada}, A.~R., {Veverka}, J., {Wagner},
  R., \& {West}, R. 2005, Science, 307, 1237

\bibitem[{{Porco} {et~al.}(2004){Porco}, {West}, {Squyres}, {McEwen}, {Thomas},
  {Murray}, {Del Genio}, {Ingersoll}, {Johnson}, {Neukum}, {Veverka}, {Dones},
  {Brahic}, {Burns}, {Haemmerle}, {Knowles}, {Dawson}, {Roatsch}, {Beurle}, \&
  {Owen}}]{Porco2004}
{Porco}, C.~C., {West}, R.~A., {Squyres}, S., {McEwen}, A., {Thomas}, P.,
  {Murray}, C.~D., {Del Genio}, A., {Ingersoll}, A.~P., {Johnson}, T.~V.,
  {Neukum}, G., {Veverka}, J., {Dones}, L., {Brahic}, A., {Burns}, J.~A.,
  {Haemmerle}, V., {Knowles}, B., {Dawson}, D., {Roatsch}, T., {Beurle}, K., \&
  {Owen}, W. 2004, \ssr, 115, 363

\bibitem[{{Simonelli} {et~al.}(1999){Simonelli}, {Kay}, {Adinolfi}, {Veverka},
  {Thomas}, \& {Helfenstein}}]{Simonelli1999}
{Simonelli}, D.~P., {Kay}, J., {Adinolfi}, D., {Veverka}, J., {Thomas}, P.~C.,
  \& {Helfenstein}, P. 1999, \icarus, 138, 249

\end{thebibliography}

%\include{table1}

%\begin{table}[h]
%   \caption[Relevant Table Title]{Relevant caption. \label{tab:relevant}}
%\begin{tabular}{ccc}
%      Chip & $Z$ (mag.) & $C$ \\ \hline
%      00 & 27.48 & -0.040 \\
%      01 & 27.81 & -0.050 \\
%\end{tabular}
%\end{table}

\newpage

\begin{deluxetable}{c d{3.2} d{3.1} d{3.1} d{3.1} d{3.1} d{3.1} d{6.1} d{3.2}  d{3.2}  d{3.2} d{3.2}  d{1.3}  d{3.1}}
%\begin{deluxetable}{l c c c c c c c c c c c c c}
\tablecaption{The geometry of all images for which a successful geometry could be determined. \label{tab:geometry}}
\floattable
\tabletypesize{\scriptsize}
\rotate
\tablehead{\colhead{Image} & \colhead{Res} & \colhead{C $\alpha$} & \colhead{C $\delta$}  & \colhead{C $\theta$} & \colhead{Sun $\alpha$}  & \colhead{Sun $\delta$}  & \colhead{C-P dist.} & \colhead{V X} & \colhead{V Y} & \colhead{N X} & \colhead{N Y} & \colhead{Vel.} & \colhead{Vel Angle}\\
\colhead{ } & \colhead{(km)} & \colhead{$(^\circ)$} & \colhead{$(^\circ)$} & \colhead{$(^\circ)$} & \colhead{$(^\circ)$} & \colhead{$(^\circ)$} & \colhead{ (km) } & \colhead{ (km) } &  \colhead{ (km) } &  \colhead{ (km) }  &  \colhead{ (km) } &  \colhead{ (km/s) } &  \colhead{ $(^\circ)$ }  }
\startdata
\multicolumn{14}{c}{Low Resolution}\\
\hline 
1465649433\_1 & 86.7 & 228.8 & -19.9 & 5.8 & 321.5 & -12.2 & 173491.5 & 237.64 & 280.53 & 634.36 & 509.53 & 0.011 & 314.5  \\
1465649746\_1 & 85.9 & 226.2 & -19.9 & 5.1 & 318.9 & -12.2 & 171865.4 & 234.83 & 275.08 & 626.96 & 508.32 & 0.069 & 341.6  \\
1465649834\_1 & 85.5 & 224.9 & -19.9 & 27.3 & 317.6 & -12.2 & 171052.3 & 233.51 & 268.79 & 575.50 & 484.98 & 0.051 & 319.9  \\
1465649979\_1 & 85.1 & 223.6 & -19.9 & 6.5 & 316.3 & -12.2 & 170239.3 & 251.39 & 282.30 & 574.88 & 499.87 & 0.222 & 304.0  \\
1465650070\_1 & 84.7 & 222.3 & -19.9 & 4.1 & 315.0 & -12.2 & 169426.2 & 235.18 & 265.06 & 572.01 & 502.83 & 0.480 & 253.4  \\
1465650234\_1 & 84.1 & 220.4 & -19.9 & 336.2 & 313.0 & -12.2 & 168206.6 & 242.61 & 258.65 & 569.60 & 488.44 & 0.077 & 263.8  \\
1465650432\_1 & 83.5 & 218.5 & -19.9 & 338.7 & 311.1 & -12.2 & 166987.1 & 294.82 & 261.62 & 643.98 & 457.69 & 0.062 & 248.5  \\
1465650745\_1 & 82.5 & 215.2 & -19.9 & 11.1 & 307.9 & -12.2 & 164954.5 & 209.01 & 252.31 & 542.13 & 449.40 & 0.219 & 326.9  \\
1465650834\_1 & 82.1 & 214.0 & -19.9 & 9.5 & 306.6 & -12.2 & 164141.4 & 287.31 & 247.61 & 629.67 & 467.70 & 0.227 & 291.3  \\
1465651001\_1 & 81.7 & 212.7 & -19.9 & 12.5 & 305.3 & -12.2 & 163328.4 & 289.89 & 261.22 & 628.01 & 463.72 & 0.305 & 247.6  \\
1465651857\_1 & 78.6 & 203.0 & -19.9 & 9.0 & 295.6 & -12.2 & 157230.6 & 264.19 & 330.12 & 742.71 & 602.22 & 0.132 & 327.8  \\
1465700253\_2 & 85.6 & 218.6 & 24.8 & 14.6 & 132.8 & -12.2 & 171291.1 & 338.98 & 291.06 & 1133.19 & 328.53 & 0.011 & 5.3  \\
1465700713\_1 & 86.9 & 214.7 & 24.8 & 18.4 & 129.0 & -12.2 & 173730.1 & 345.15 & 309.59 & 1175.00 & 333.68 & 0.201 & 248.5  \\
1465700837\_1 & 87.5 & 212.8 & 24.8 & 16.5 & 127.0 & -12.2 & 174949.6 & 339.89 & 308.17 & 1143.69 & 326.59 & 0.371 & 270.6  \\
\multicolumn{14}{c}{Medium Resolution}\\
\hline 
1465661929\_1 & 44.5 & 95.0 & -19.8 & 348.5 & 186.9 & -12.2 & 88941.4 & 191.21 & 151.68 & 462.08 & 316.27 & 0.061 & 316.4  \\
1465662167\_1 & 43.7 & 92.4 & -19.8 & 341.9 & 184.3 & -12.2 & 87315.8 & 175.49 & 148.95 & 424.73 & 299.01 & 0.135 & 259.0  \\
1465662631\_1 & 42.2 & 87.9 & -19.8 & 350.7 & 179.8 & -12.2 & 84470.9 & 164.61 & 145.83 & 420.46 & 297.76 & 0.329 & 310.4  \\
1465664774\_1 & 34.9 & 65.0 & -19.8 & 345.8 & 156.5 & -12.2 & 69841.5 & 127.83 & 54.96 & 351.51 & 134.01 & 0.149 & 152.3  \\
1465665036\_1 & 34.1 & 62.4 & -19.8 & 330.2 & 153.9 & -12.2 & 68216.2 & 139.50 & 56.56 & 336.76 & 135.08 & 0.085 & 218.9  \\
1465665440\_1 & 32.7 & 58.0 & -19.7 & 341.1 & 149.4 & -12.2 & 65372.0 & 65.65 & 78.24 & 251.50 & 201.28 & 0.018 & 256.3  \\
1465665563\_1 & 32.3 & 56.7 & -19.7 & 339.7 & 148.1 & -12.2 & 64559.4 & 160.35 & 80.58 & 352.10 & 191.66 & 0.042 & 283.4  \\
1465665771\_1 & 31.7 & 54.8 & -19.7 & 0.7 & 146.1 & -12.2 & 63340.5 & 121.93 & 147.14 & 330.74 & 293.23 & 0.598 & 272.1  \\
\multicolumn{14}{c}{High Resolution}\\
\hline 
1465669068\_1 & 20.3 & 19.8 & -19.6 & 344.9 & 109.9 & -12.2 & 40597.3 & 177.34 & 124.46 & 456.47 & 240.97 & 0.515 & 88.8  \\
1465669741\_1 & 18.1 & 13.1 & -19.5 & 345.2 & 102.8 & -12.2 & 36133.4 & 29.63 & 11.56 & 146.38 & 71.87 & 0.085 & 282.1  \\
1465669944\_1 & 17.5 & 11.3 & -19.5 & 335.2 & 100.8 & -12.2 & 34916.4 & 163.47 & 29.84 & 278.77 & 76.56 & 0.129 & 300.0  \\
1465670212\_1 & 16.2 & 7.7 & -19.4 & 2.5 & 97.0 & -12.2 & 32482.9 & 193.01 & 313.06 & 160.58 & 144.58 & 0.452 & 277.7  \\
1465670650\_1 & 15.0 & 4.2 & -19.4 & 3.5 & 93.1 & -12.2 & 30050.3 & 171.67 & 427.40 & 150.56 & 267.49 & 0.111 & 288.1  \\
1465671285\_1 & 13.0 & 358.4 & -19.3 & 351.5 & 86.6 & -12.2 & 25998.9 & 186.80 & 42.48 & 170.88 & 37.63 & 0.205 & 158.6  \\
1465671448\_1 & 12.4 & 356.7 & -19.2 & 6.6 & 84.7 & -12.2 & 24784.4 & 7.54 & 87.84 & -2.01 & 86.74 & 0.565 & 252.4  \\
1465671822\_1 & 11.2 & 353.4 & -19.2 & 1.8 & 80.8 & -12.2 & 22357.0 & 160.42 & 166.51 & 138.82 & 93.65 & 0.423 & 274.5  \\
1465672161\_1 & 10.0 & 350.3 & -19.0 & 350.0 & 76.9 & -12.2 & 19932.6 & 103.01 & 100.92 & 85.80 & 27.27 & 0.149 & 88.2  \\
1465673600\_1 & 5.2 & 341.2 & -17.9 & 325.2 & 61.4 & -12.2 & 10305.6 & 67.79 & -44.15 & 74.66 & -66.69 & 0.173 & 310.3  \\
1465677443\_1 & 8.1 & 98.0 & 25.6 & 71.2 & 20.0 & -12.2 & 16149.5 & 69.98 & 65.53 & 136.23 & 124.61 & 0.166 & 284.4  \\
1465677670\_1 & 8.9 & 96.2 & 25.5 & 52.9 & 17.4 & -12.2 & 17760.2 & 82.08 & 79.43 & 131.94 & 144.11 & 0.262 & 289.3  \\
1465678419\_1 & 11.7 & 89.0 & 25.4 & 57.0 & 8.3 & -12.2 & 23414.2 & 149.44 & 304.52 & 192.06 & 140.61 & 0.309 & 165.2  \\
1465678911\_1 & 13.1 & 85.1 & 25.3 & 72.8 & 3.8 & -12.2 & 26247.3 & 134.89 & 332.74 & 148.61 & 152.18 & 0.245 & 127.5  \\
1465679413\_1 & 14.7 & 80.5 & 25.2 & 84.0 & 358.6 & -12.2 & 29488.1 & 136.31 & 289.02 & 156.90 & 125.14 & 0.028 & 293.0  \\
1465679675\_1 & 15.6 & 78.2 & 25.2 & 81.4 & 356.0 & -12.2 & 31109.4 & 104.41 & 301.79 & 134.42 & 139.35 & 0.215 & 260.1  \\
1465679932\_1 & 16.4 & 75.8 & 25.2 & 79.7 & 353.4 & -12.2 & 32731.2 & -13.12 & 191.41 & 291.15 & 241.92 & 0.305 & 249.3  \\
1465680977\_2 & 21.0 & 61.8 & 25.1 & 85.3 & 338.6 & -12.2 & 42062.9 & 228.57 & 200.53 & 585.70 & 235.91 & 0.375 & 102.2  \\
\enddata
\tablecomments{C and P refer to the Cassini spacecraft and Phoebe. X and Y are the horizontal and vertical shifts of Phoebe from the bottom left corner of the VIMS visible (V) and NIR (N) detectors, respectively. $\alpha$, $\beta$, and $\theta$ correspond to longitude, latitude, and azimuth respectively. The reported sub-Cassini and sub-Solar $\alpha$ and $\beta$ values, and the C-P distance are the adjusted from the JPL Horizons ephemeris as described in Section 2.}
\end{deluxetable}

\end{document}